\documentclass[sigconf, nonacm, review=false]{acmart}

\AtBeginDocument{%
  }

\setcopyright{acmlicensed}
\copyrightyear{2024}
\acmYear{2024}
\acmDOI{XXXXXXX.XXXXXXX}

\begin{document}

\title{Information Flows for Athletes' Health and Performance Data}

\author{Brad Stenger}
\affiliation{%
  \institution{University of Vermont \\ Department of Computer Science}
  \city{Burlington}
  \state{VT}
  \country{USA}
}

\author{Yuanyuan Feng}
\affiliation{%
  \institution{University of Vermont \\ Department of Computer Science}
  \city{Burlington}
  \state{VT}
  \country{USA}
}

\renewcommand{\shortauthors}{Stenger, Feng}

\begin{abstract}
Increasing numbers of athletes and sports teams use data collection technologies to improve athletic development and athlete health with the goal of improving competitive performance. Personal data privacy is managed but it is not always a priority for the coaches who are in charge of athletes. There is a pressing need to investigate what are appropriate information flows as described by contextual integrity\cite{nissenbaum_privacy_2010,shvartzshnaider_learning_2016} for these data technologies and these use cases. We propose two main types of information flows for athletes' health and performance data---team-centric and athlete-centric---designed to characterize data used for the collective and individual physical, psychological and social development of athletes. We also present a scenario for applying differential privacy to athletes' data and propose two new information flows---research-centric and community-centric---which envision larger-scale, more collaborative sharing of athletes' data in the future.
\end{abstract}




\maketitle

\section{Introduction}

Wearable technology and computer vision have made sports a data-rich environment. These are technologies that collect data for each athlete in a team setting. The data collectors track changes of position (of the body or a limb), capture biometric information (like heart rate) or log (automatically or manually) relevent, variable biological processes (nutrition, sleep/recovery, fluctuating biomarkers). As the level of competition improves and the stakes increase, more technology comes into play, and more data is collected for analysis. Professional athletes incorporate the most data into their training and competition\cite{australian_academy_of_science_getting_2022}. Inter-collegiate athletes frequently use health and performance data for training and competition\cite{luczak_state---art_2020,robell_pac-12_2023}, and increasing numbers of high school and teenage athletes collect athletes' data\cite{jovanovic_examining_2021}.

Athletes generally can refuse to participate data collection, but often choose not to \cite{kolovson_personal_2020}. Social norms around a team-first ethic in sports encourage personal sacrifice, and with sacrifice comes trust, for coaches and teammates\cite{malloy_investigating_2023}. Teams are hubs for athletes' data collection but are not the sole endpoint for organizing and applying athletes' data. Schools, municipalities and small businesses oversee sports teams. Leagues oversee competing teams. Sports associations oversee sports at local, regional, national and international levels. A hierarchy that starts at local schools and sports leagues extends to powerful international and professional organizations that administer the Olympics, FIFA soccer, NFL football, among many professional sports groups.

As technologies for collecting athletes' health and performance data continues to improve the situation becomes more urgent to account for the information flows of this data. Team-centric and athlete-centric information flows fit the parameters of contextual integrity and should provide a substantial, if still basic, understanding. The initial contextual understanding sets up further investigation of more complex phenomena: teammates and relationship dynamics, commercial vendor partners, oversight organizations, and norms for data sharing among teams.

\section{Information Flows}

Competitive athletes have in recent years grown increasingly reliant on data that records ongoing changes to their health and to physical and biomechanical changes related to their athletic performance\cite{australian_academy_of_science_getting_2022}. The data is used to guide athletic training for an individual's athletic development and to gauge team preparation for competition. Most of the data is collected in a team context. Even athletes who participate in individual sports like running or swimming do their training in groups overseen by coaches and their use of data also fits this paradigm. 

\subsection{Team-centric Information Flow}

The information flows for athletes' data are mostly team-centric\cite{australian_academy_of_science_getting_2022}. In CI-terms if the sender, the subject and the information type are athletes and their health and performance data, then the primary recipient is the coach in charge. Other team recipients (assistant coaches, medical, and administrative staff) exchange athletes' information collected from the entire team with the coach in charge. Athletes are recipients, but usually at their coaches' discretion. Competitive advantage and athletes' development are the two transmission principles, and in many contexts the two go hand-in-hand.

An example of inappropriate information flow is when coaches use player data to highlight players' inadequate effort or performance without permission or transparency\cite{stenger_injury_2013}. Kolovson studied the power dynamics in the coach-athlete relationship and the use of player tracking data and identified privacy abuse as an unchecked control that coaches can apply without clear benefit to either the team or a privacy-compromised player\cite{kolovson_personal_2020}.  

Appropriate information flow is consistent with sports coaching competence. In the U.S. coaching certification and training requirements vary from sport to sport, and from place to place. Unlike the U.S. Canada's government established the Coaching Association of Canada (CAC) in 1970 to educate and certify sports coaches at all levels of competition. The CAC lists the things that coaches should aim to undertake\cite{coaching_association_of_canada_coaching_nodate}:
\begin{itemize}
\item{Encourage participants to be active and have fun}
\item{Plan purposeful practices and create engaging activities}
\item{Help participants develop sport skills}
\item{Provide constructive criticism to help participants improve performance}
\item{Manage problems by making ethical decisions}
\item{Enable safe participation through a safe environment}
\item{Teach participants how to respect themselves, others, and their sport.}
\end{itemize}

\subsection{Athlete-centric Information Flows}

Athlete-centric information flows exist. Athletes can, and sometimes do exert agency to either misrepresent or circumvent team-centric information flow. An injured or overworked athlete rarely answers questions about their own health or exertion truthfully. To do so might jeopardize status within the team\cite{minett_peer_2022}.

Athletes can also use personal health and performance technology on their own, and they can enlist personal coaches outside of the team context. The athletes themselves are the primary recipients of athlete-centric information flow. The use of technology to collect data in team-centric information flow takes away athletes' agency by limiting an athletes available pathways to development within the team\cite{kolovson_using_2024}.

To understand the complexity of information flows for athletes' data it helps to understand the biopsychosocial model of athlete development\cite{brown_mind_2020}. Important measurable "bio" aspects of the model include: metabolism, biomechanics, location tracking, biometrics, biomarkers, nutrition, recovery/sleep and athletes' gender differences. "Psycho" aspects include: well-being, anxiety, stress and confidence. "Social" aspects include: sharing, sacrifice, collaboration, and relationships with coaches, peer athletes and adversaries. Realizing athletic potential is what occurs when all three facets contribute to and improve athletes' overall development.

\subsection{Teammates}

Mutual interest and mutual benefit between athlete and team is significant for athlete development, and they are dynamic elements that affect the interpersonal athlete-coach and athlete-teammate relationships\cite{martin_developing_2017}. Friendships based on shared interests and close collaboration are inevitable (and beneficial) but it creates a fuzzy boundary where teammates can be recipients in either team-centric or athlete-centric information flows. The biopsychosocial model puts social development on par with physical and psychological development, and navigating interpersonal relationships on a team is also crucial for competitive success.

Team-centric and athletic-centric information flows are dynamic. Interpersonal relationships change constantly among athletes, coaches and teammates. Athletes' development can also change rapidly and unpredictably as they age and they improve (or decline). 

\subsection{Injuries}

Injuries significantly change the day to day activity of athletes. The normal routine of team practices and competitions is replaced with recovery, rehabilitation and therapy. Injured athletes spend more time away from than with coaches and teammates until they have completed sports medicine return-to-play and return-to-perform protocols which have their own technological and data practices\cite{luczak_state---art_2020} (and information flows). Athletes recovering from injuries have altered relationships with coaches and teammates that can change the overall social context. Interested care professionals may also be CI recipients, and it can be important to distinguish if the care is on the behalf of the athlete or the team.

%

\subsection{Products and Vendors}

Teams and athletes use commercial technology products to gather and manage athletes' health and performance data. Only the highest levels of competition have resources to customize commercial technology and/or build their own solutions\cite{stenger_sports_2017}. One exception: Some teams make extensive use of spreadsheets instead of using commercial athlete management software. Some athletic technology vendors have a CI sender role to play, with company tech support augmenting the expertise of coaches who will help athletes to use the equipment in a data collection context. Some athletic technology vendors have a CI recipient role as part of a team's athlete data management solution. The active roles of vendors are functional, in support of coaches' evidence-based decision-making. But these commercial products have constraints that arise from product design and the feature sets these products offer.

Commercial products for athletes data are the primary safeguards for maintaining athletes' privacy. Most of these solutions store teams' athletes' data in the cloud. These products also collect signed waiver forms for privacy policy agreements. Teams vary in the amount of attention that they pay to these agreements. Coaches are, in general, thankful give responsibility for athlete privacy to vendors or to a team staffperson who is paid to manage and operate technology.

\subsection{Sports Organizational Hierarchy and Information No-Flows}

Sports operates with a hierarchy of leagues and other oversight administrators that govern what is allowable by teams with regard to competition, and they also have the authority to act in the interests of athletes when teams fail to do so. Some college and professional sports leagues have taken an active role in collecting and researching athletes' health and performance data. The Pacific 12 sports conference collected athletes data across mens and womens sports for research purposes\cite{robell_pac-12_2023}. The NFL has an active program with Amazon to collect and manage football players' running speeds and tackle impacts in order to improve equipment requirements for player safety\cite{ghasem_player_2021}.

The team-centric transmission principles of competitive advantage and athlete development can inhibit athletes' data sharing\cite{lamberts_evolution_2022}. Teams avoid sharing information with competitors unless it has been mandated by their league. The focus on realizing athletes' potential by teams and athletes willfully blinds them to the biopsychosocial limitations that athletes possess. These are information no-flows where norms preclude data sharing regardless of privacy.

\section{Scenario: Athletes' Differentially Private Data Information Flows}

Differential privacy (DP) is privacy-enhancing technology that introduces noise to dataset queries in order to render personal information indistinguishable. The noise sacrifices some of the dataset's accuracy but not enough to completely discount the dataset's utility. Accounting for the amount of noise lets a data provider manage the tradeoff between privacy and utility Recent work by Benthall and Cummings highlights benefits from combining the two research areas\cite{benthall_integrating_2024} and recommends adding a "transmission property" parameter to CI that captures essential information about DP that affects context. The relationship between CI and DP is a two-way street. Cummings has also written about the need for greater contextual information in DP deployments.

Athletes' health and performance data is a good candidate for DP deployment. Differential privacy might diffuse (or eliminate) teams' disincentive to share athletes' data based on competitive advantage. The information flows for DP-implemented athletes' data are likely to be different from team-centric or athlete-centric information flows.

\subsection{Differential Privacy}

Differential privacy (DP)~\cite{dwork2014algorithmic}  is a formal privacy definition designed to allow statistical analysis while protecting information about individuals. Differentially private analyses, often called \emph{mechanisms}, typically add random noise to analysis results in order to achieve privacy. Formally, two datasets $D, D' \in \mathcal{D}$ are called \emph{neighboring datasets} if they differ in one person's data, and a mechanism $\mathcal{M}$ satisfies $(\epsilon, \delta)$-DP if for all neighboring datasets $D$ and $D'$ and sets of outcomes $S$:
\vspace*{-7pt}
\[ \Pr[\mathcal{M}(D) \in S] \leq e^\epsilon \Pr[\mathcal{M}(D') \in S] + \delta \]

\vspace*{-7pt} \noindent The $\epsilon$ parameter is the \emph{privacy parameter} or \emph{privacy budget}; a smaller $\epsilon$ results in stronger privacy, while a larger $\epsilon$ results in weaker privacy.

In the U.S. the National Institute of Standards and Technology (NIST) has drafted guidelines for deploying differential privacy\cite{nist_nist_2023}. NIST describes the privacy afforded by DP with a "privacy pyramid." (Figure 1) The top level of the pyramid, epsilon and "unit of privacy" are direct measures of users' privacy guarantees. The middle tier are factors that can undermine a differential privacy guarantee. The bottom tier are data security fundamentals. Each component in the pyramid depends on the components below it.

\begin{figure}
    \centering
    \includegraphics[width=0.5\linewidth]{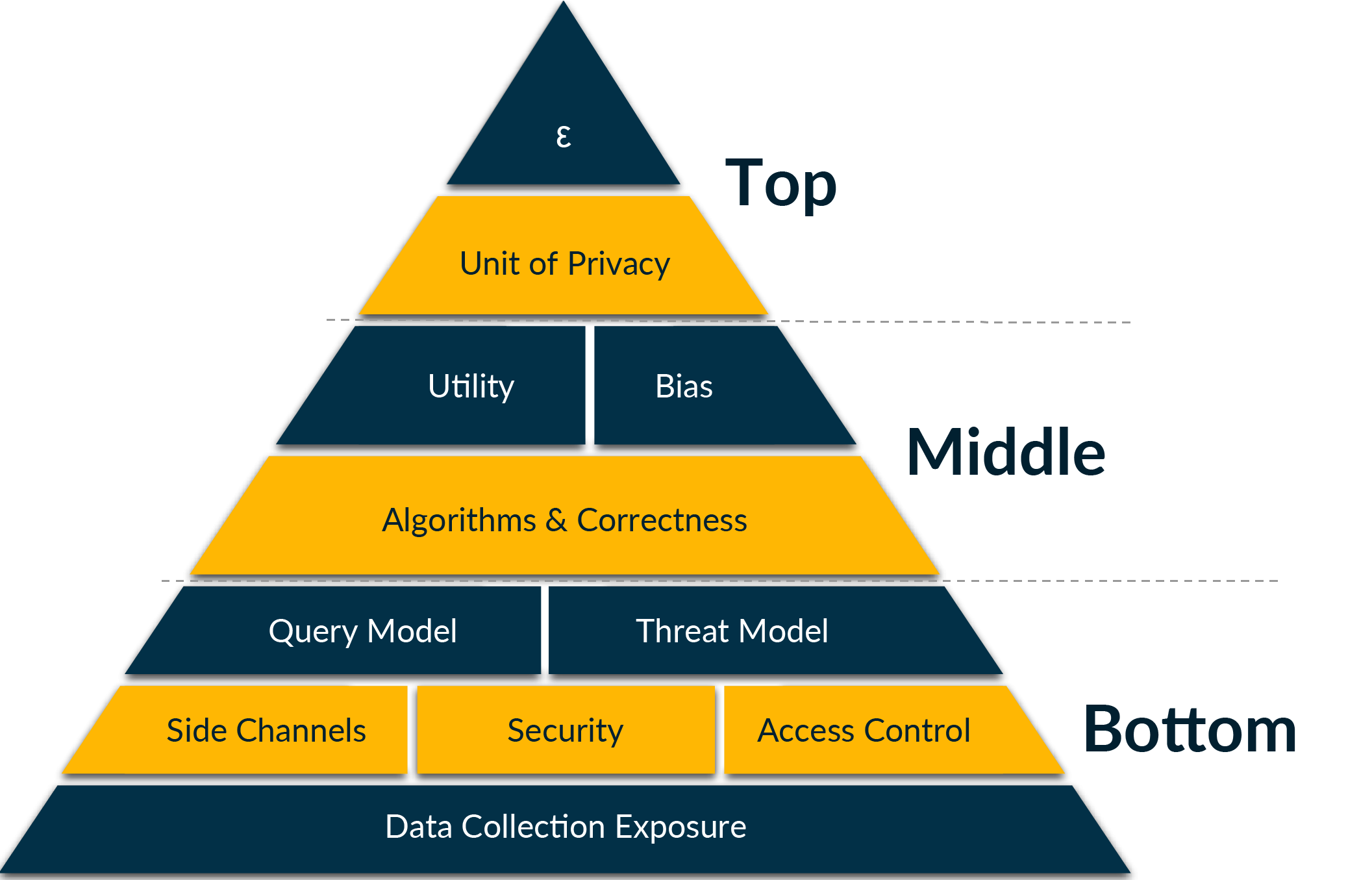}
    \caption{NIST Privacy Pyramid - effective differential privacy implementation require suitable DP parameters (top), careful data structures (middle), and good info-sec fundamentals (bottom)}
    \label{fig:enter-label}
\end{figure}
\subsection{CI and DP}

Benthall and Cummings' recent paper on Integrating Differential Privacy and Contextual Integrity\cite{benthall_integrating_2024} builds on the CI framework to incorporate aspects of DP that describe the privacy implementation in a social privacy context. They recommend adding a new CI parameter called "transmission properties" to describe the algorithmic and technical properties of an information flow. Expanding the scope of CI enables interpretations of more complex data flows and bridges CI arguments to related applied privacy research.

Cummings and colleagues also argue in a published comment\cite{cummings_comment_2024} to the NIST guidelines that DP stands to benefit from CI's understanding of privacy in its social and normative context with regard to the collection, analysis and reporting of data. The authors also point out a growing need for instructive case studies of DP deployments (whether successful or unsuccessful).

\subsection{Research-centric Information Flow} 

A new information flow based on differential privacy for athletes' data could be research-centric, helping to answer important questions about challenges to athletes' biopsychosocial development. CI recipients would be interested health, social science and technology investigators. CI subjects take athletes' data into the most important problems and questions related to today's youth sports and athlete development: How can young athletes maximize their enjoyment of sports? How can all athletes minimize injuries? 

The enormous variability of athletic body types and biomechanics means that datasets larger than what one team can assemble are necessary for solving difficult problems related to athletes' health and performance. The historical reluctance by teams to share data can be diminished with differential privacy, assuming DP can conceal athletes' identities and that doing so preserves competitive advantage for teams.

The challenge of creating athletes' data that implements DP is the potential for physical signatures in biometric and biomechanical data that identify specific athletes. As athletes progress to higher competitive levels their development is often the result of unique physical characteristics. Those characteristics may be impossible to conceal using DP, at least not with an acceptable level of data utility. The transmission properties of a research-centric information flow would identify these essential tradeoffs so that data providers can knowingly agree to provide public datasets using DP, and so that data users can understand how and where data accuracy has been sacrificed.

\subsection{Community-centric Information Flow}
Another new information flow based on differential privacy for athletes' data could be community-centric, fostering buy-in and consensus among sports' diverse stakeholders. The starting point for a community could be academic, geographic or ideological, and not necessarily sport-specific, but where common interests drive CI subject and recipient. CI recipients would be the stakeholder community and the transmission principle would be the costs or benefits of the situation's outcome. A general example might occur when a league wants to construct a new shared facility, and every team needs to agree on it.

Research-centric and community-centric information flows for athletes' data are not mutually exclusive and could well be reinforcing, where community supports research, and research organizes communities.

\section{Conclusion}

Team-centric and athlete-centric information flows are promising starting points for a thorough investigation of the contextual integrity of athletes' health and performance data. Research-centric and community-centric information flows are potentially valuable results of a scenario where differential privacy is applied to athletes' health and performance data.

%

%

\begin{acks}
Primary support for this work comes from the National Science Foundation under award \#2348294. Any opinions, findings and conclusions or recommendations expressed in this material are those of the author(s) and do not necessarily reflect those of the National Science Foundation.
\end{acks}

\bibliographystyle{ACM-Reference-Format}
\bibliography{zotero}

\end{document}